# TRANSLATIONAL KNOWLEDGE MAP OF COVID-19

**Cesar Aguado-Cortés and Victor M. Castaño***

Universidad Nacional Autónoma de México

Av. Universidad 3000, Cd. Universitaria

Ciudad de México, 04510 MEXICO

*corresponding author: vmcastano@unam.mx

**Abstract**

A translational knowledge map of COVID-19, based on the analysis of scientific papers and networks citation concurrence of terms and keywords of the terms: covid-19, 2019-ncov and sars-cov-2 in leading databases (MEDLINE, web of Science and Scopus), was constructed. Some fields of the research on covid-19 are connected together, differing in structure, content and evolution.

**Introduction**

COVID-19 epidemic is an emergent infectious disease caused by respiratory syndrome coronavirus 2 and has been named SARS-CoV-2 (formerly 2019-nCoV) and COVID-19 disease (1-8). Coronaviruses were first described in 1965 (1) by Tyrell and Bynoe, who cultured the viruses from patients with what was thought common colds symptoms (2,3). The term "Coronavirus," which described the characteristic microscopic morphology of these viruses, was officially coined in 1968 (3). SARS CoV-2 belongs to the B-lineage of beta-coronaviruses and is closely related to the SARS-CoV virus. Currently, the suspicion of infection with 2019-nCoV requires two elements as indicators of the case: presence of fever and symptoms of respiratory disease. SARS-CoV-2 apparently managed to make its transition from animals to humans at the seafood market in Wuhan, China (4,5). Wuhan, is the same place where one of the largest biosafety level 4 (BSL-4) Asia´s laboratories is located, where European, Canadian and American experts collaborate. The initial clinical sign of COVID-19-related disease that allowed detection was pneumonia. Observations so far suggest an incubation average period of five days.

The literature on the medical, virology and epidemiology aspects of the COVID-19 emergency are beginning to abound, revealing important pieces of data towards the understanding and possible solutions of this global problem (4-8). However, very scarce studies are yet available on the analysis of the conceptual mapping of coronaviruses, in terms of the relationships among all the data produced in the last 25 years in this area. Accordingly, this present investigation shows an analysis of the published documents of COVID-19 in reliable sources, including Medline, Web of Science and Scopus, and the results of the mentioned bases were integrated into a comprehensive bibliometric analysis, including not only basic and applied research, but also how this knowledge begins to involve intellectual property (11).

**Methodology**

The software package Mendeley, a recognized references manager, was employed to collect 547 items of different databases such as Medline, Web of Science and Scopus. To map the knowledge translation of COVID-19, we used the statistical indicator R and graphics, networks and histograms through the software package "Bibliometrix". This methodology has proven very insightful to the knowledge structure of complex diseases cases such as cancer (9) and ebola (10). Figure 1 shows schematically the analytical methodology we have employed (9-10).

Data was extracted from different scientific and medical databases to analyze the different and relevant concepts that revolve around COVID-19, including 2019-nCoV and SARS-CoV-2. Then, a data framework from different articles was created, along with a structure of the involved knowledge, that allows to understand the relationships among the different variables. Thus, results from the different Medline, Web of Science and Scopus databases can be appreciated unified and not separated from each other, as there is the case of the few studies available, which present relevant data, although disconnected and somehow incoherent. Therefore, to estimate the relative risk of transmission of COVID-19, we considered the most relevant scientific databases to estimate the risk of using a single analysis.

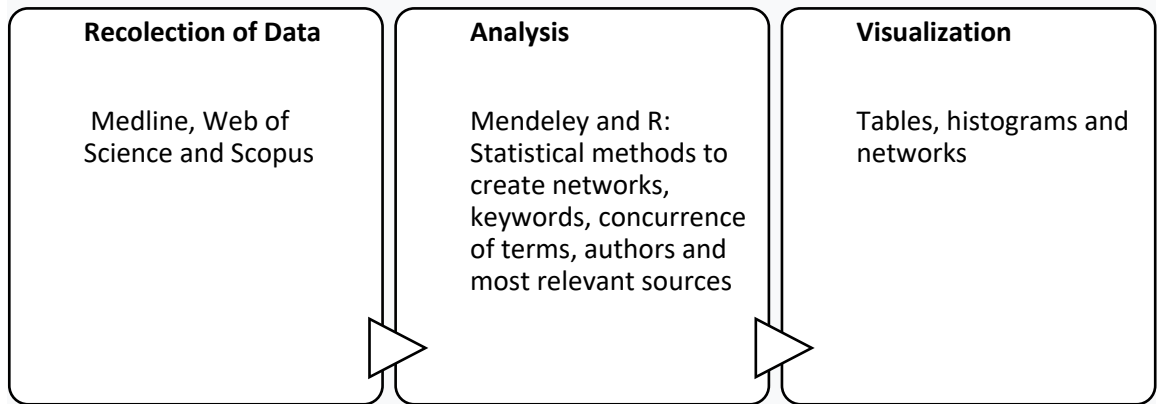

Figure 1. Block diagram of the analysis

**Results and discussion**

We analyzed 547 articles from the information contained in different scientific databases, based on the key terms, so the statistical applications determine the behavior of the variables, as summarized in Table 1.

| Description | Results |
|---|---|
| Documents | 547 |
| Sources | 184 |
| Keywords | 160 |
| Period | 2003 - 2020 |
| Authors | 2198 |
| Author Appearances | 3315 |
| Authors of single-authored documents | 70 |
| Authors of multi-authored documents | 2128 |
| Single-authored documents | 121 |
| Documents per Author | 0.249 |
| Authors per Document | 4.02 |
| Co-Authors per Documents | 6.06 |
| Collaboration Index | 5 |

Table 1.-Contents of the Data Bases

To create a conceptual structure of COVID-19 one requires not only to properly identify the relevant concepts involved, but also their time and geographical evolution. This allows to identify that 2020 has been the year with the highest number of scientific articles, as shown in Table 2. This would be an obvious result, given the current worldwide concern on this disease. However, the time evolution of the knowledge related to associated keywords reveals interesting facts.

| Year | Articles |
|---|---|
| 2003 | 2 |
| 2004 | 1 |
| 2006 | 1 |
| 2011 | 1 |
| 2019 | 1 |
| 2020 | 522 |

Table 2. Yearly scientific production

The origin of the references is very important, to identify the interest and influence of various geographical locations, as well as the disciplines which were first attracted to this phenomenon, as observed in Table 1, which provides a first clue that, despite representing an area of scientific interest and public health danges, coronavirus has been basically a set of isolated efforts, localized in few authors and institutions who focused on this subject, with a relatively low Collaboration Index.  Certainly, there are are few single-autored papers, but the great percentage correspond to teams working  consistently in the area for nearly 20 years. The  most relevant and most cited sources are summarized in Table 3.

| Source | Articles |
|---|---|
| LANCET (LONDON ENGLAND) | 34 |
| JOURNAL OF MEDICAL VIROLOGY | 32 |
| CHINESE JOURNAL OF TUBERCULOSIS AND RESPIRATORY DISEASES | 18 |
| BMJ (CLINICAL RESEARCH ED.) | 16 |
| RADIOLOGY | 16 |
| JOURNAL OF CLINICAL MEDICINE | 15 |
| THE NEW ENGLAND JOURNAL OF MEDICINE | 13 |
| TRAVEL MEDICINE AND INFECTIOUS DISEASE | 13 |
| EURO SURVEILLANCE : BULLETIN EUROPEEN SUR LES MALADIES TRANSMISSIBLES | 12 |
| THE LANCET. INFECTIOUS DISEASES | 12 |
| INTENSIVE CARE MEDICINE | 10 |
| THE JOURNAL OF INFECTION | 10 |

Table 3. Most relevant bibliographic sources

Figure 2 allows to better visualize the time evolution of the knowledge generated around coronavirus. Interestingly, and despite that the first reports, published in top journals, date back to the 1960´s (1-3), only in the very recent past a clear increase in the production of knowledge can be appreciated, coinciding with the first claim of novelty in 2014, which has resulted in a patent granted in 2018 (11).

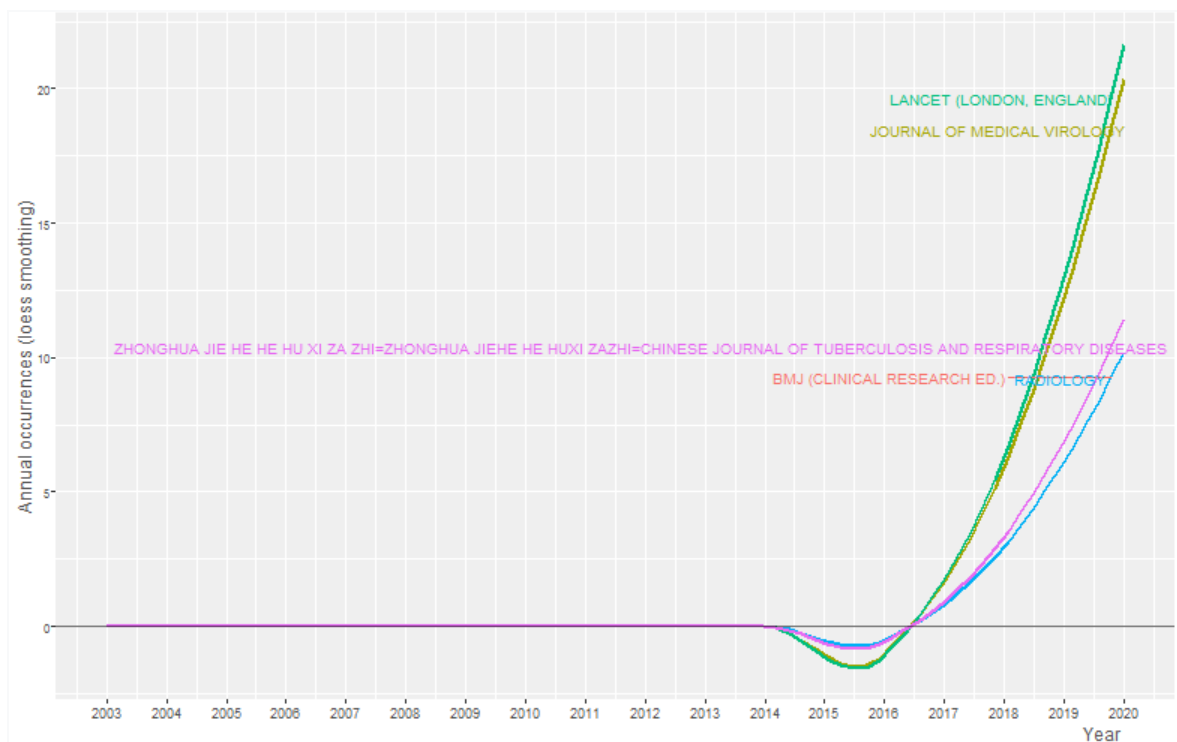

Figure 2. Time evolution of sources

It is also interesting to analyze the most relevant authors in what now is known as COVID-19, not only by names, but also from their professional associations. Thence, we find Y. Wang with 21 publications, X. Li with 17 posts, W. Wang with 15 articles, Y. Yang with 15 papers, among others, as shown in Figure 3.

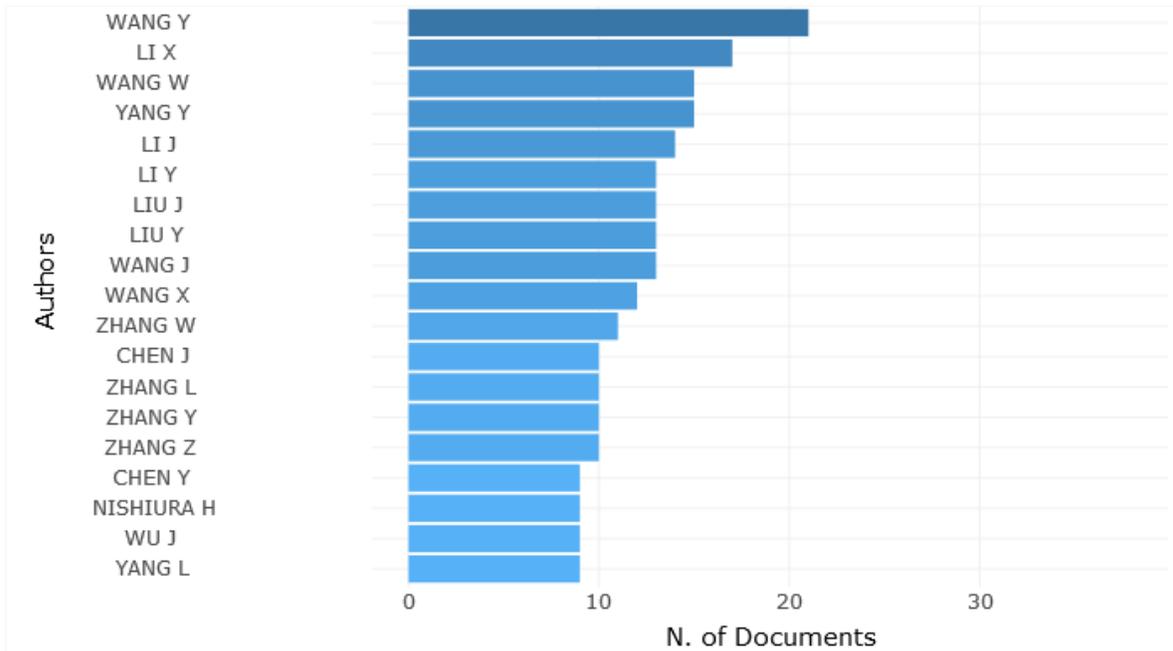

Figure 3. Production by main authors

Then, the constructed database was filtered by keyword layers: Abstracts, Titles and Author's keywords, as shown in Figures 4, 5 and 6. Notice that, despite that China was officially the origin of COVID-19, in the keyword layers the term "China" is not among the most relevant ones, except when comes to authors. "Coronavirus", "Patients" and "Humans" are the leading terms in the literature. The term "COVID" is also among the most important indicators.

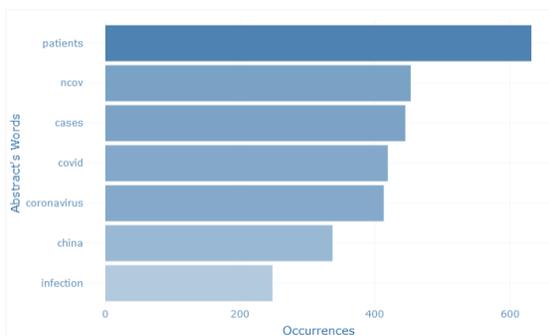

Figure 4. Layer Abstracts

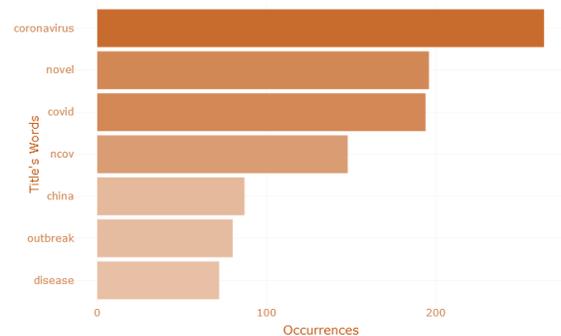

Figure 5. Layer Titles

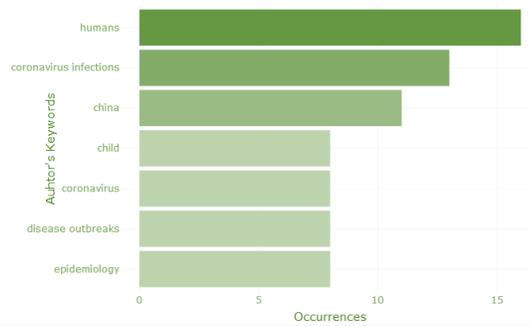

Figure 6. Layer Author's keywords.

| Words | Occurrences |
|---:|---:|
| patients | 632 |
| outbreak | 80 |
| novel | 196 |
| ncov | 601 |
| infection | 248 |
| humans | 16 |
| epidemiology | 8 |
| disease outbreaks | 8 |
| disease | 72 |
| covid | 613 |
| coronavirus infections | 13 |
| coronavirus | 677 |
| coronavirus | 8 |
| china | 435 |
| cases | 445 |
| TOTAL | 4,052 |

Table 4. Ocurrence of keywords

Figure 7 shows the concurrence network of the set of articles analyzed.

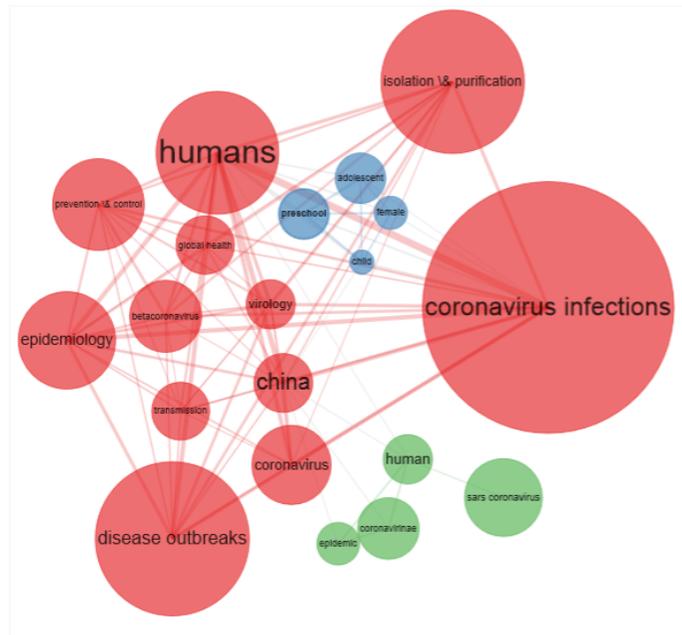

Figure 7. Concurrence network of articles

When Centrality is analyzed, Table 5, the term "Humans" is the most important one, followed by "China" nd "Coronavirus infections", with the rest of the nodes having very little contribution. This allows to quantify the influence of those particular nodes within the network of knowledge formed by all the articles and patents.

| Term | Centrality |
|---|---|
| humans | 47.7859514864345 |
| china | 38.2912056482265 |
| human | 19.4310344827586 |
| coronavirus infections | 17.6666184907487 |
| child | 0.98803659394792 |
| epidemiology | 0.51042434528946 |
| isolation & purification | 0.35951366609041 |
| virology | 0.35951366609041 |
| betacoronavirus | 0.19368131868131 |
| prevention & control | 0.19368131868131 |
| disease outbreaks | 0.15254237288135 |
| transmission | 0.06779661016949 |

Table 5. Centrality of terms

As it could be expected, COVID-19 involves a wide variety of terms, concepts, references, citations and actions which, at first thought, could seem impossibly to clasify or organize in a coherent way, given the multidisciplinary character of the problem. However, as it can be appreciated in Figure 8, it is possible to structure all the information available into few conceptual clusters, which can be, in turn, summarized by basically 3 concepts, namely "China", "Coronavirus" and "Disease outbrake". This has been consistent in the last 17 years,which allows to understand why it not should be surprising that a "Coronavirus outbrake took place in China".

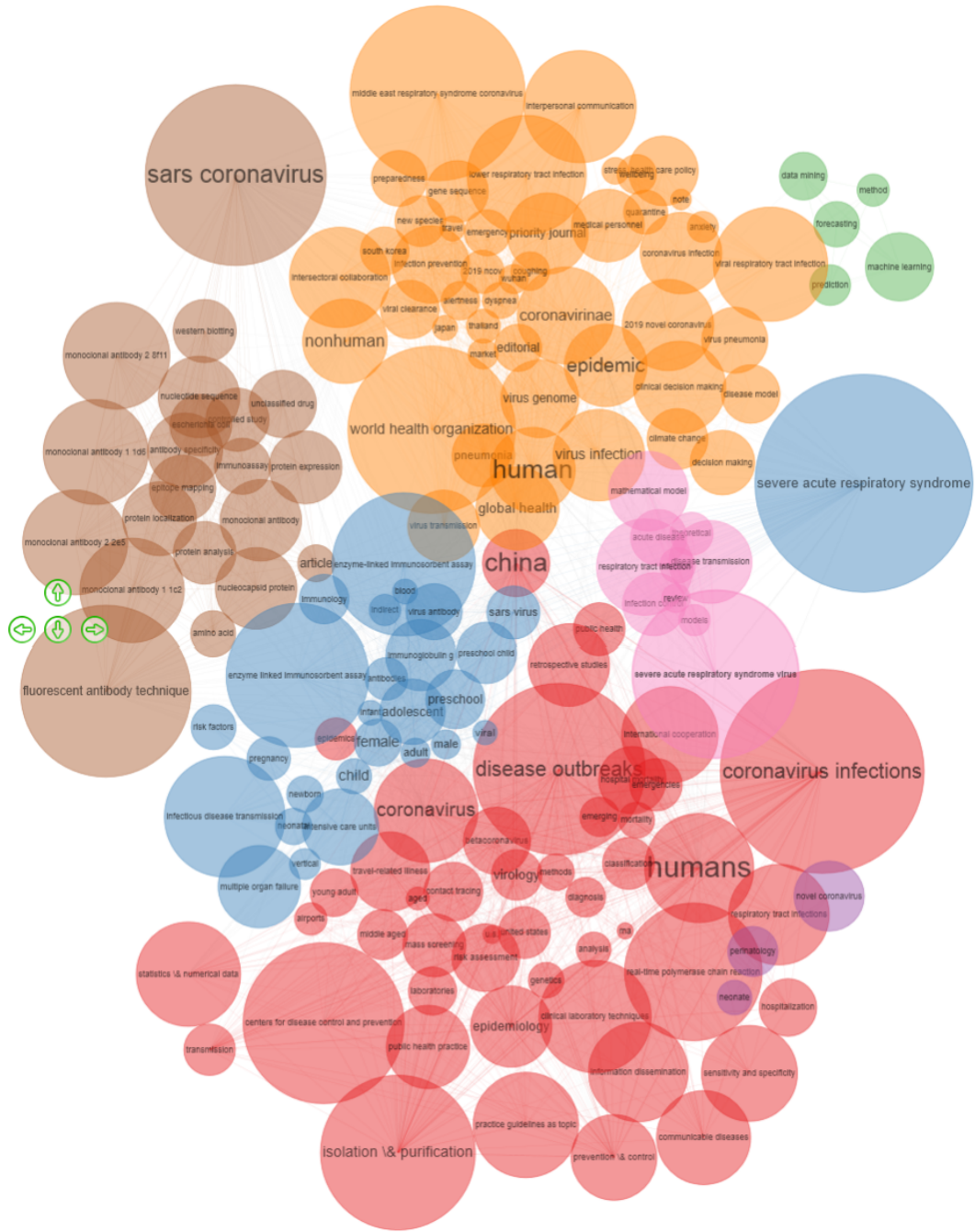

Figure 8. Conceptual structure of COVID-19

Finally, if we redraw that data of Figure 8 into a conceptual map, Figure 9, which corresponds to the actual Conceptual Map of COVID-19, which allows to identify the modularity of the nodes that represent the entire database built from the scientific information available.

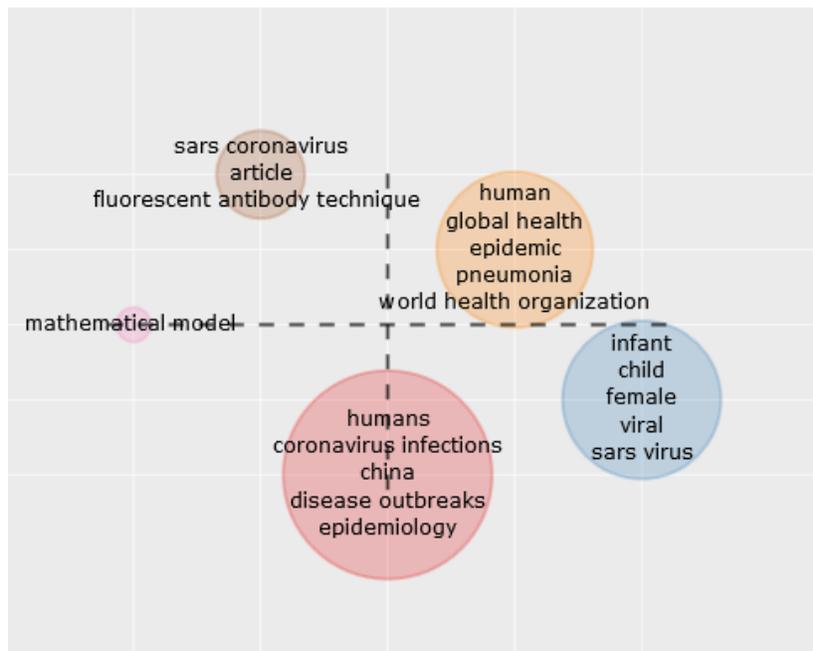

Figure 9. Conceptual Map of COVID-19

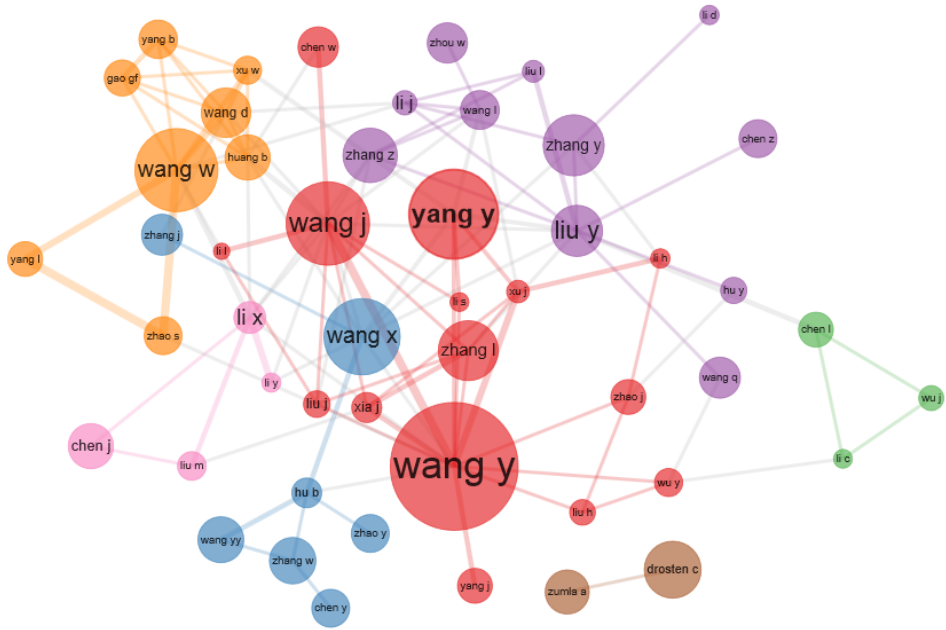

Figure 9. Collaboration of the authors.

A knowledge cluster is centered at the origin (SARS coronavirus), while the other focus on the causes (epidemic) symptoms-free (infant and child), creation of scenarios (mathematical models) and on the fibal consequences (coronavirus infections). A translation process of the knowledge in research through a set of items that link the infection by covid-19 with pneumonia via high degree of infection and transmission in humans, was found.

**Concluding remarks**

A translational knowledge map of COVID-19 was constructed from the available literature data on articles, reports and patents, aiming to provide a relevant vision of the current stage and an overview of the knowledge structure of COVID-19. Furthermore, this methodology has the potential to become a useful assessment tool for monitoring the evolution of knowledge on various emerging disease such as COVID-19. Finally, it should be mentioned that, despite that the fundamental scientific fact is that COVID-19 outbreak began affecting China, and therefore these terms are at the center of the conceptual structure, there exist less visible nodes identified by terms such as: "war", "biotechnology", "fear", "economic" and "economy", which appear related to patents (11).